\documentclass[twocolumn]{aastex61}
\usepackage{graphicx}
\usepackage{subfigure}
\usepackage{amsmath}
\usepackage{epsfig}

\newcommand{\refsec}[1]{Section~\ref{#1}}
\newcommand{\reffig}[1]{Fig.~\ref{#1}}

\newcommand{\IllustrisTNG}{\texttt{IllustrisTNG}}
\shorttitle{Kinematic Decomposition Using auto-GMM}
\shortauthors{Du et al.}

\shorttitle{auto-GMM}
\shortauthors{Du et al.}

\begin{document}
\title{Identifying kinematic structures in simulated galaxies using unsupervised machine learning}

\correspondingauthor{Min Du}
\email{dumin@pku.edu.cn}

\author{Min Du}
\affil{Kavli Institute for Astronomy and Astrophysics, Peking University, Beijing 100871, China}

\author{Luis C. Ho}
\affiliation{Kavli Institute for Astronomy and Astrophysics, Peking University, Beijing 100871, China}
\affiliation{Department of Astronomy, School of Physics, Peking University, Beijing 100871, China}

\author{Dongyao Zhao}
\affil{Kavli Institute for Astronomy and Astrophysics, Peking University, Beijing 100871, China}

\author{Jingjing Shi}
\affil{Kavli Institute for Astronomy and Astrophysics, Peking University, Beijing 100871, China}

\author{Victor P. Debattista}
\affiliation{Jeremiah Horrocks Institute, University of Central Lancashire, Preston PR1 2HE, UK}

\author{Lars Hernquist}

\affiliation{Harvard--Smithsonian Center for Astrophysics, 60 Garden Street, Cambridge, MA 02138, USA}

\author{Dylan Nelson}
\affiliation{Max-Planck-Institut f$\ddot{u}$r Astrophysik, Karl-Schwarzschild-Str. 1, 85741 Garching, Germany}

\begin{abstract}

Galaxies host a wide array of internal stellar components, which need to be 
decomposed accurately in order to understand their formation and evolution.  
While significant progress has been made with recent integral-field 
spectroscopic surveys of nearby galaxies, much can be learned from analyzing 
the large sets of realistic galaxies now available through state-of-the-art
hydrodynamical cosmological simulations. We present an unsupervised machine 
learning algorithm based on Gaussian mixture models, named auto-GMM, to 
isolate intrinsic structures in simulated galaxies based on their kinematic phase space.  
For each galaxy, the number of Gaussian components allowed by the data is determined 
through a modified Bayesian information criterion. We test our method by applying it to 
prototype galaxies selected from the cosmological simulation \IllustrisTNG. Our method can 
effectively decompose most galactic structures. The intrinsic structures 
of simulated galaxies can be inferred statistically by non-human supervised 
identification of galaxy structures. We successfully 
identify four kinds of intrinsic structures: cold disks, warm disks, bulges, and halos.
Our method fails for barred galaxies because of the complex kinematics of particles moving on bar orbits.

\end{abstract}
\keywords{galaxies: fundamental parameters --- galaxies: kinematics and dynamics --- galaxies: structure --- methods: numerical}

\section{Introduction}

The Hubble (1926) sequence, the most widely used system of morphological 
classification of galaxies \citep{Sandage1981}, has served as a powerful 
framework for understanding galaxy evolution. Notwithstanding their myriad 
complexities, at the most fundamental level galaxies are principally 
distinguished by two dominant structural/dynamical components: a fast-rotating,
flattened disk and a pressure-supported, spheroidal bulge. The relative light 
fraction of these two components, traditionally determined through photometric 
decomposition \citep[e.g.,][]{Peng2002, Mendez-Abreu2008, Erwin2015, Gao2019}, 
establishes the galaxy type. Early-type 
galaxies are pure spheroids or bulge-dominated disk systems, whereas late-type 
galaxies are increasingly disk-dominated and even bulgeless. The processes 
that build up bulges and disks underlie the physical basis of the Hubble 
sequence.

Galaxies often comprise additional structures. For 
example, many nearby galaxies have a thick disk, which is both older and more 
metal-poor with respect to the thin disk \citep{Dalcanton&Bernstein2002, 
Yoachim2006, Comeron2011, Comeron2014, Elmegreen2017}. Meanwhile, the 
morphology of bulges comes in more than one flavor, ranging from highly 
spherically symmetric to flat \citep{Andredakis&Sanders1994, Andredakis1995, 
Courteau1996, Mendez-Abreu2010}. Classical bulges are dynamically hot and largely featureless, 
likely the end-products of galaxy major mergers \citep{Toomre1977}. The more 
flattened, rotationally supported pseudo bulges, an outgrowth of internal 
secular evolution, generally coexist with complex central structures \citep[e.g.,][]{Kormendy&Kennicutt2004, Erwin2004}. 
The Milky Way is a prototypical spiral that has several components, including a thin and 
a thick disk, a boxy/peanut-shaped bulge, a bar, a stellar halo, and a 
nuclear star cluster \citep[see review by][]{Bland-Hawthorn&Gerhard2016}. 
Boxy/peanut-shaped bulges are generally considered as the vertically thickened part of bars.
Whether the Milky Way has a classical bulge is still uncertain, but it is 
unlikely to have a massive one \citep{Shen2010, Debattista2017}. The rich 
diversity of substructures observed among nearby galaxies imprints the 
formation and evolutionary history of galaxies. Accurate recognition and 
decomposition of these underlying substructures is essential.

The rapid development of integral-field spectroscopy has enabled 
galaxies to be classified by their internal kinematics 
\citep[e.g.,][]{Emsellem2007, Emsellem2011, Cappellari2011a, Cappellari2011b}. 
Early-type galaxies can be classified into slow and fast rotators \citep[see 
the review of][and references therein]{Cappellari2016}, with fast-rotator 
early types forming a parallel sequence to spiral galaxies. \citet{Zhu2018b} 
made the first attempt to decompose observed galaxies based on their kinematics.
Using the orbit-superposition Schwarzschild method 
\citep[e.g.,][]{Schwarzschild1979, Valluri2004, vandenBosch2008}, they reconstructed stellar orbits for 
galaxies in the CALIFA survey \citep{Sanchez2012}, decomposing them into cold, 
warm, and hot components \citep{Zhu2018a, Zhu2018c}. However, given the limited
information that can be extracted from spectra, it is still very difficult to 
decompose observed galaxies in detail. 

Numerical simulations are powerful tools for studying the formation and 
evolution of galaxy structures. 
In recent years, significant progress has been made in modelling star formation and 
stellar feedback, leading to increasingly realistic galaxies with reasonable 
bulge-to-disk ratios \citep{Agertz2011, Guedes2011, Aumer2013, Stinson2013b, Marinacci2014,
Roskar2014, Murante2015, Colin2016, Grand2017}. 
The increase of simulation resolution has enabled us to generate galaxies with multiple 
structures that go much beyond the basic bulge+disk system, including
vertical structures of disks 
\citep{Brook2012b, Ma2017, Navarro2018, Obreja2018b}, 
stellar halos \citep{Cooper2010, Tissera2013, Pillepich2014, Elias2018, Monachesi2019}, 
pseudo bulges \citep{Okamoto2013, Guedes2013}, and bars \citep{Algorry2017, Peschken2019}.

Large-scale hydrodynamical cosmological simulations provide the opportunity 
to investigate the
statistical properties of galaxies evolving in a fully cosmological context.
Recent advances include Illustris \citep{Vogelsberger2014a, Vogelsberger2014b,
Genel2014}, EAGLE \citep{Schaye2015, Crain2015}, and Horizon-AGN 
\citep{Dubois2016}. The IllustrisTNG simulations \citep{Nelson2018a, Nelson2019a, Marinacci2018, Naiman2018, Pillepich2018b, Pillepich2019, Springel2018} can reproduce galaxies that successfully 
emulate plausible visual morphologies, thanks to an updated galaxy physics model 
\citep{Weinberger2017, Pillepich2018a}. The optical morphologies of galaxies 
in the TNG100 run (the highest-resolution version currently available at $z=0$) are 
in good agreement with observations of nearby galaxies 
\citep{Rodriguez-Gomez2019, Huertas-Company2019}. The realism 
of the mock galaxies inspires confidence that the latest simulations can be 
used for detailed statistical study. With the aid of numerical simulations in which 
information is known in all six dimensions of phase space, we can 
investigate the intrinsic properties of galaxy structures, as well as track 
their formation physics and evolutionary history. 

The structures of simulated galaxies can be identified through the kinematic 
properties of their constituent stars. \citet{Abadi2003b} proposed a 
circularity parameter $\epsilon=J_z/J_c$, the ratio of the azimuthal angular 
momentum $J_z$ and the maximum angular momentum $J_c$ having the same binding 
energy $E$, that can separate effectively the spheroidal component from the 
disky component. In order to characterize different components in detail, 
\citet{Domenech-Moral2012} further introduced into consideration the binding 
energy $E$ and the non-azimuthal angular momentum vector ${\bf J_p}={\bf J - J_z}$,
where ${\bf J}$ is the total angular momentum vector of the stellar particle. 
These parameters identify the clustering of particles in kinematic phase space 
that corresponds to intrinsic structures of a galaxy. \citet{Obreja2016, 
Obreja2018a} replaced the $k$-means clustering algorithm used in 
\citet{Domenech-Moral2012} with an unsupervised machine learning algorithm, 
the Gaussian mixture model (GMM). The use of GMM reduces the errors caused by 
the mixtures of different structures via soft assignment of stars. 

This study extends the application of GMM and develops a method to decompose 
simulated galaxies automatically. We test the method by applying it to five prototype galaxies
from the TNG100 simulation \citep[now also publicly available; 
see][]{Nelson2019a}, and we discuss prospects for forthcoming applications 
using larger samples of simulated galaxies.

\section{Dynamical Decomposition Method}
\label{method}

Stars belonging to the same physical structure naturally cluster in their 
kinematic phase space. The method of \citet{Domenech-Moral2012} and 
\citet{Obreja2018a} offers a promising framework to decompose the complex 
internal structures of galaxies. Their method uses three-dimensional Gaussian 
distributions to represent structures identified through their kinematics. 
However, a few limitations still affect its application:
 
\begin{itemize}

\item The number of structures is determined artificially. This not only 
opens the possibility of human bias, but also renders impractical 
implementation to large samples of galaxies from cosmological simulations.

\item Real galaxy structures may not follow simple single-Gaussian 
distributions. The distribution function of disks, possibly all structures 
in galaxies, are not single Gaussians. A given structure may be 
composed of more complex distribution functions. Moreover, any realistic,
dynamic, evolving system inevitably contains some degree of finer 
substructure.

\end{itemize}

An automated method, such as the unsupervised machine learning nature of GMM, 
is needed to explore this problem. To mitigate human bias, the number of Gaussian
components should be inferred directly from data, and all galaxies must be 
treated with the same standard. The fits should be sufficiently detailed to 
resolve significant structures in galaxies, allowing the same structure to host 
more than one Gaussian component if necessary. Our fully automated methodology, 
auto-GMM, complies with all these requirements and is able to identify multiple 
kinematic structures. We do not ascribe any physical 
significance to each individual component, postponing to a later stage the 
interpretation of the components/sub-components and their association with 
known, observed structures.

\subsection{3D Kinematic Phase Space}
\label{phasespace}

For TNG100, we load the positions and velocities of all particles 
(including dark matter, gas, and star) within a selected subhalo. 
Then, the code of \citet{Obreja2018a} is adopted to calculate the three-dimensional (3D) kinematic phase 
space of $j_z/j_c$, $j_p/j_c$, and $e/|e|_{\rm max}$ of stars, which will be used as 
inputs to GMM. The quantities $j_z$, $j_p$, $j_c$, and $e$ are the specific 
$J_z$, $J_p$, $J_c$, and $E$, respectively. 
The origin of the coordinate coincides with the galaxy center, which is defined as 
the minimum of the gravitational potential. 
The $z$-axis of the galaxy is oriented perpendicularly to the outer disk. The average angular momentum vector 
is calculated by stars whose radii are between $2.1$ kpc (3 times the softening radius of stars) and 
0.1 times the virial radius. Then, the azimuthal term of the 
angular momentum $j_z$ can be easily decomposed from $j_p$. In order to 
estimate $j_c$ and $e$, the code recalculates the gravitational potential of 
the halo, under the assumption that the halo is isolated. This assumption is 
generally well satisfied, unless the galaxy is undergoing significant 
accretion, which is fairly rare at low redshifts. All of the dark matter, 
stellar, and gaseous masses are included to recalculate the gravitational 
potential. The quantity $|e|_{\rm max}$
is the absolute value of the energy of the most bound stellar particle in the 
halo. Thus, $e/|e|_{\rm max}$ describes how tightly bound or centrally 
concentrated a particle is. It is a dimensionless parameter that gives a 
typical value across all galaxy masses. 
Only particles with $j_z/j_c \in [-1.5, 1.5]$, $j_p/j_c \in [0, 1.5]$, 
and $e\in [-1, 0]$ are considered in dynamical decomposition, 
consistent with the criteria used in \citet{Obreja2018a}. 
These criteria reject interlopers that are particles with clearly 
different kinematics from the galaxy.

\subsection{Gaussian Mixture Models}
\label{GMM}

\begin{figure*}[htbp]
\begin{center}
\includegraphics[width=1.0\textwidth]{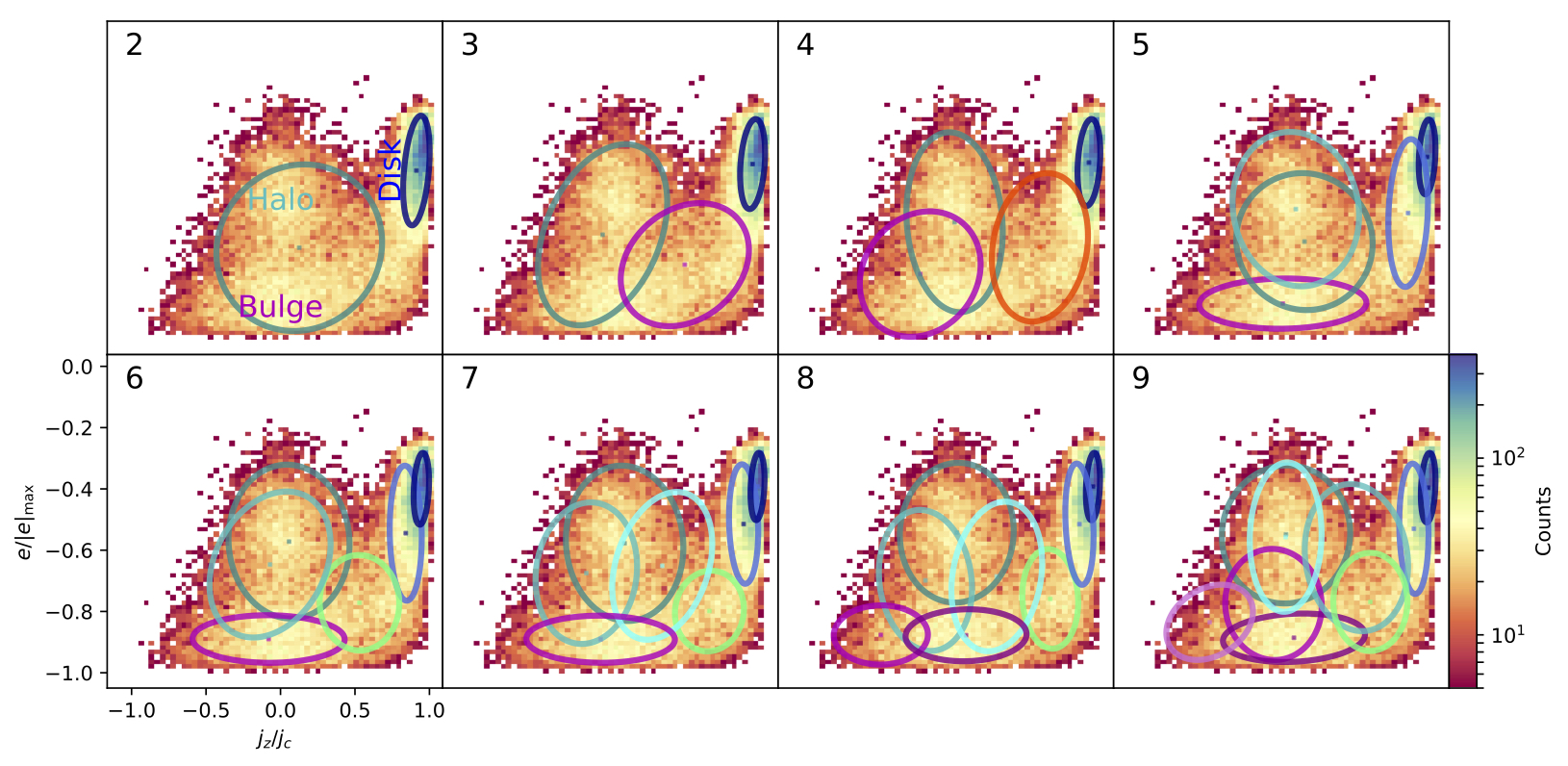}
\caption{The distribution of $j_z/j_c$ versus $e/|e|_{\rm max}$ of a typical galaxy. Three components, likely corresponding to a disk, a bulge, and a halo, are visible. The color bar indicates the number of stellar particles in each bin. We fit the distribution of particles with increasing numbers of Gaussian components, from 2 to 9, as labelled in the upper-left corner of each panel. The overlaid ellipses represent $63\%$ confidence regions of the Gaussian components found by the GMM fits. The fit improves by adding more components. Both the bulge (magenta ellipses) and disky components ($j_z/j_c>0.7$; blue ellipses) are well fitted using $n_c = 5-8$. With the increase of $n_c$, the halo breaks up into multiple substructures (cyan ellipses).}
\label{fig:jze}
\end{center}
\end{figure*}

\begin{figure*}[htbp]
\begin{center}
\includegraphics[width=1.0\textwidth]{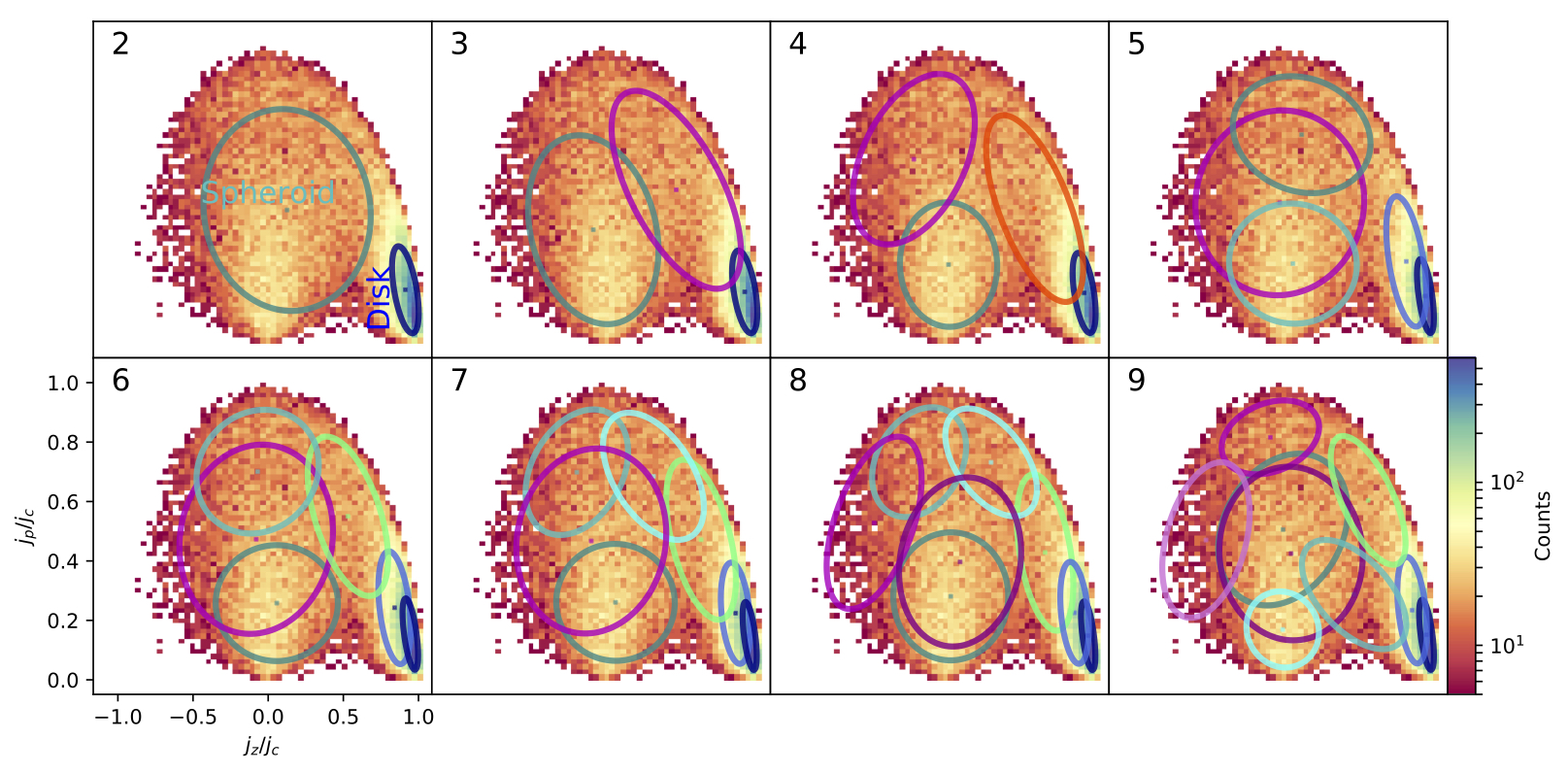}
\caption{The distribution of $j_z/j_c$ versus $j_p/j_c$ of the same galaxy shown in Figure~\ref{fig:jze}. A spheroidal and a disky component are clearly visible. The color bar indicates the number of stellar particles in each bin. We fit the distribution of particles with increasing numbers of Gaussian components, from 2 to 9, as labelled in the upper-left corner of each panel. The Gaussian components found by GMM are represented by ellipses, using the same color scheme as in Figure~\ref{fig:jze}. }
\label{fig:jzjp}
\end{center}
\end{figure*}

Unsupervised machine learning algorithms can be used to cluster data points 
into different groups. The \texttt{PYTHON} language \citep{scikit-learn} offers 
several clustering methods. As suggested by \citet{Obreja2018a}, GMM is 
suitable for finding structures in the kinematic phase space of $j_z/j_c$, 
$j_p/j_c$, and $e/|e|_{\rm max}$. In the updated {\tt scikit-learn} package, 
the old GMM module is replaced by the {\tt GaussianMixture} module. Each 
Gaussian component is a triaxial ellipsoid in kinematic phase space. In order 
to maximize the likelihood in the parameter space, an expectation-maximization 
algorithm iterates until the default criterion 
is satisfied, returning a matrix of probabilities. Each data point has a 
probability array of how likely it is that it belongs to a certain component. 

As an example, Figures \ref{fig:jze} and \ref{fig:jzjp} illustrate the
kinematic phase space of a galaxy with distinct spheroidal and disky 
structures. Three components are clearly visible in the $j_z/j_c$ versus 
$e/|e|_{\rm max}$ diagram (\reffig{fig:jze}), namely a compact and slow-rotating spheroid or bulge, 
a diffuse and slow-rotating spheroid or halo, and a fast-rotating disk (details about the identification 
of kinematic and morphological structures are provided in \refsec{sec:intri}).
By contrast, only two components are clearly seen in the 
$j_z/j_c$ versus $j_p/j_c$ diagram (\reffig{fig:jzjp}). This is because both 
spheroidal components, dominated by random motions, have a wide range of 
$j_p/j_c$.

We fit the kinematic phase space with GMM, varying the number of Gaussian 
components $n_c$ from 2 to 9. In each case, the fit is performed 10 times with 
different initializations by setting the keyword {\tt n\_init$=10$}. We 
emphasize that running enough initializations is very important to obtain a 
stable fit. All initial parameters are generated with the $k$-means algorithm. 

Figures \ref{fig:jze} and \ref{fig:jzjp} use colored ellipses to represent the 
$63\%$ confidence ellipse of each Gaussian distribution obtained by the GMM 
fit. A disky and a spheroidal component can be roughly represented by setting 
$n_c=2$ (top-left panel), but the kinematics of the galaxy are apparently more 
complex than such a simple, conventional bulge+disk decomposition. As expected, the 
fit improves by adding more components, but the data clearly become overfit 
when $n_c\geq9$. The kinematics of this galaxy are well reproduced with $n_c 
= 5-8$. Both the bulge (magenta ellipses) and disky components ($j_z/j_c>0.7$;
blue ellipses) are well fitted, while the halo breaks up into multiple 
substructures (cyan ellipses) with increasing $n_c$. The halo and bulge 
components show no distinct separation in the $j_z/j_c$ versus $j_p/j_c$
diagram. Thus, components identified purely in the $j_z/j_c$ versus $j_p/j_c$ 
plane are not as robust as those decomposed in the $j_z/j_c$ versus 
$e/|e|_{\rm max}$ plane.

\subsection{Bayesian Information Criterion}
\label{BIC}

Instead of artificially choosing $n_c$, we derive a modified version of the 
Bayesian information criterion (BIC) for selecting $n_c$ in GMM. The BIC,
developed by \citet{Schwarz1978} and widely used in analysis of clustering 
data, allows the user to infer an approximate posterior distribution over the 
parameters of a Gaussian mixture distribution. Its formal definition is

\begin{equation}
{\rm BIC} = -2 n \cdot {\rm ln} (\widehat{L}) + k \cdot {\rm ln} (n), 
\end{equation}

\noindent
where ${\rm ln} (\widehat{L})$ is the average log-likelihood of a given data set, 
$n$ is the number of data points, and $k$ is the number of free parameters 
to be estimated. Because the geometry of each Gaussian distribution is fully 
relaxed by allowing a free 3D covariance matrix, GMM adds 10 extra free 
parameters (1 weight, 3 means, and 6 covariances) for each additional Gaussian 
component (i.e. $k=10n_c$). The BIC is a decreasing function of 
${\rm ln} (\widehat{L})$ and an increasing function of $k$. Hence, the second 
term on the right-hand side of Equation (1) is a penalty for the number of 
parameters introduced in the fit and serves to limit overfitting. A model 
having a smaller BIC is preferred, which implies either fewer free parameters 
or a better fit. 

The mean BIC of each data point is 

\begin{equation}
\widehat{\rm BIC}(n, n_c) = \frac{\rm BIC}{n} = -2 {\rm ln} (\widehat{L}(n_c))+ \frac{10n_c{\rm ln} (n)}{n}.
\end{equation}

\noindent
This form is more meaningful, as $\widehat{\rm BIC}$ quantifies how good a model
is for each single stellar particle. However, it is not completely independent 
of $n$. We vary $n_c$ from 2 to 15. Because $n$ is generally rather large 
($\gtrsim 10^5$), the penalty term is $\lesssim 0.01$, estimated from the case 
of $n=10^5$ and $n_c=10$. As a consequence, we cannot see a clear minimum 
$\widehat{\rm BIC}$; instead, $\widehat{\rm BIC}$ approaches an asymptotic
value that changes little for $n_c>10$. Additionally, as suggested in 
\refsec{GMM}, using more than 10 Gaussian components in the fit is not 
well motivated physically. We define 

\begin{equation}
\Delta \widehat{\rm BIC} = \widehat{\rm BIC} - \widehat{\rm BIC}_{\rm min},
\end{equation} 

\noindent
where $\widehat{\rm BIC}_{\rm min}=\sum_{n_c=11}^{15} \widehat{\rm BIC}(n_c)/5$
is the mean value of $\widehat{\rm BIC}(n_c>10)$. $\Delta \widehat{\rm BIC}$ 
of every galaxy asymptotically reaches $\sim 0$. The number of components can 
be chosen as the minimum value that satisfies $\Delta \widehat{\rm BIC}<
C_{\rm BIC}$, where $C_{\rm BIC}$ is our criterion for a reasonable GMM model. 
The choice of $C_{\rm BIC}$ will be discussed in the following section.

\begin{figure}[htbp]
\begin{center}
\includegraphics[width=0.48\textwidth]{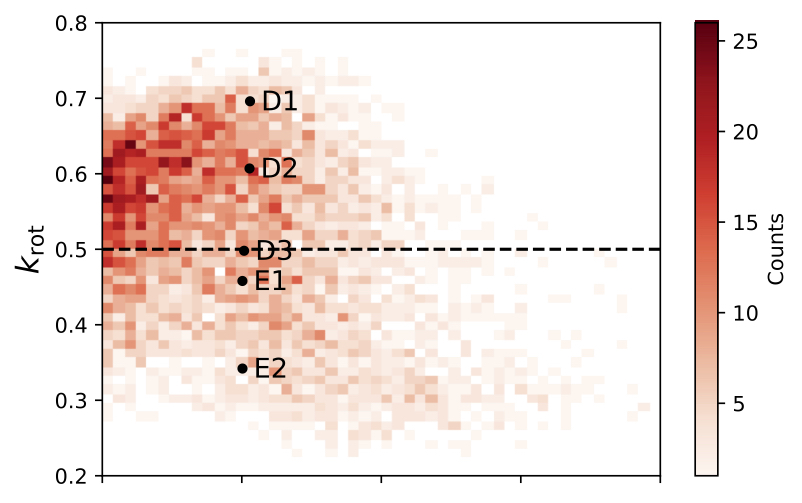}
\includegraphics[width=0.48\textwidth]{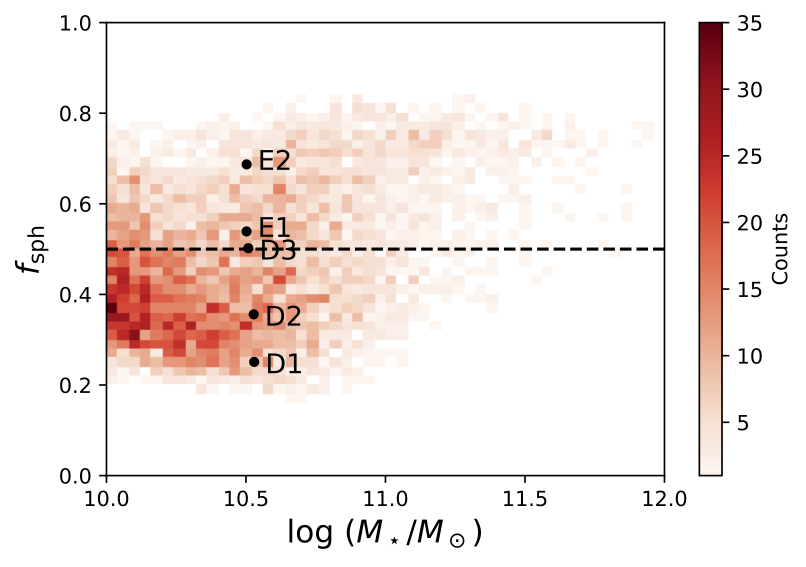}
\caption{The fraction of the kinetic energy in ordered rotation, $K_{\rm rot}$, 
of the stellar particles (top) and the fraction of the stellar mass in the spheroidal component (bottom), $f_{\rm sph}$, as a function of stellar mass, $M_\star$, for the TNG100 galaxies. Here all 6503 galaxies, including barred galaxies, are shown.  Five prototypes are marked with solid dots, classified into three disk galaxies (D1, D2, D3) and two ellipticals (E1, E2) with the criterion $K_{\rm rot}=0.5$ (dashed lines). The color bar represents the number of galaxies in each bin. A total of 2994 unbarred disk ($K_{\rm rot}\geq0.5$) galaxies are used for further analysis in Sections \ref{sec:BIC} and \ref{sec:intri}.}
\label{fig:Krot}
\end{center}
\end{figure}

Our approach of combining GMM with BIC, which we call auto-GMM, takes advantage of the unsupervised 
nature of GMM and allows $n_c$ to be inferred objectively and automatically 
from the data, with no additional assumptions imposed.

\section{Application of Auto-GMM to Prototype Galaxies from \IllustrisTNG}
\label{prototypes}

Auto-GMM allows us to decompose galaxies automatically and efficiently, 
making it a powerful tool for large data sets. In order to test the efficiency 
of this method, we apply it to prototype galaxies at redshift 0 from the
TNG100. 

\subsection{$C_{\rm BIC}$ Inferred from the \IllustrisTNG\ Galaxies}
\label{sec:BIC}


\begin{figure}[htbp]
\begin{center}
\includegraphics[width=0.48\textwidth]{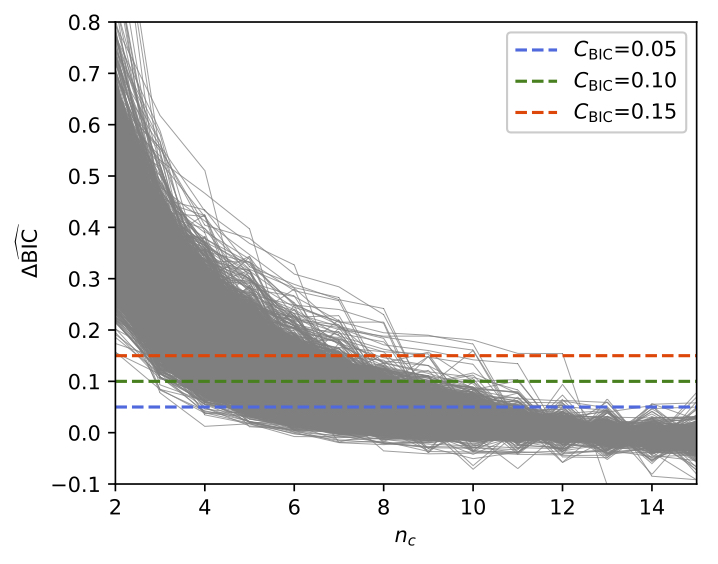}
\caption{The $\Delta \widehat{\rm BIC}$ profiles as a function of $n_c$. All unbarred galaxies of stellar mass $> 10^{10} M_\odot$ are included.
We only exclude the elliptical galaxies, which are largely dominated by random motions ($j_z/j_c<0.2$). The horizontal dashed lines 
mark different positions for the criterion $C_{\rm BIC}$ with values 0.05, 0.1, and 0.15.}
\label{fig:BICprofiles}
\end{center}
\end{figure}

\begin{figure}[htbp]
\begin{center}
\includegraphics[width=0.48\textwidth]{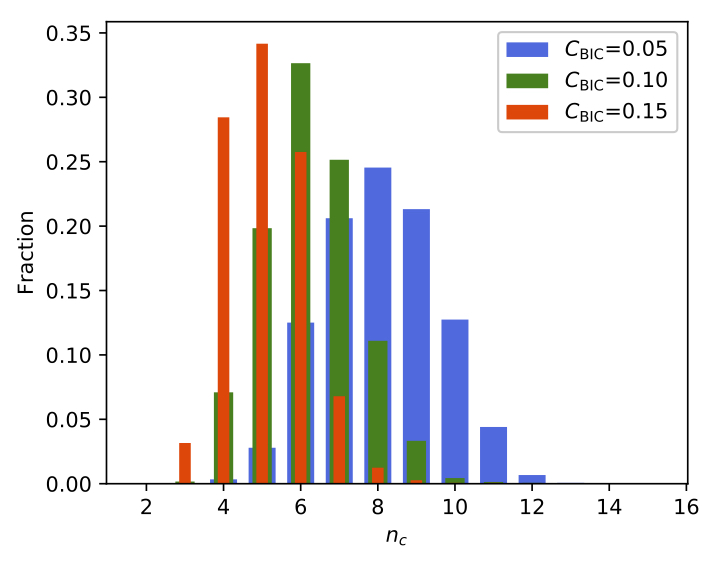}
\caption{The distribution of the number of components $n_c$ chosen by criterion $C_{\rm BIC}=0.05, 0.1$, and 0.15 for the same galaxies shown in \reffig{fig:BICprofiles}.}
\label{fig:nchist}
\end{center}
\end{figure}

\begin{figure*}[htbp]
\begin{center}
\includegraphics[width=0.98\textwidth]{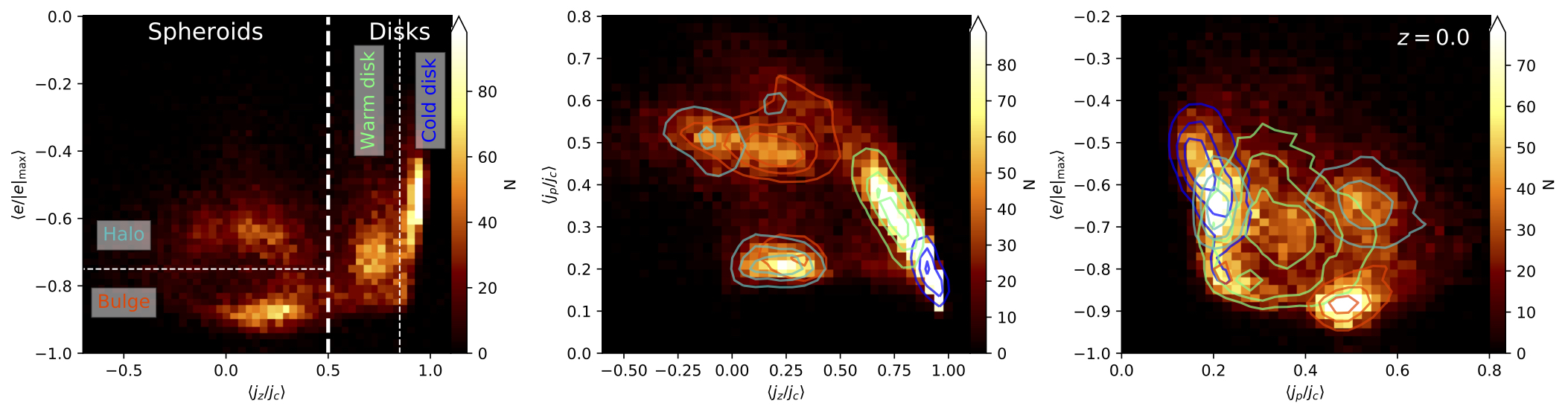}
\caption{The kinematic phase space of all components of the unbarred disk galaxies from TNG100. The quantities 
$\langle j_z/j_c \rangle$, $\langle j_p/j_c \rangle$, and $\langle e/|e|_{\rm max} \rangle$ are the mean values 
of each Gaussian component found by auto-GMM with $C_{\rm BIC}=0.1$. The color bar indicates 
the number of components in each bin. Four distinguishable clusters emerge in the diagram of 
$\langle j_z/j_c \rangle$ versus $\langle e/|e|_{\rm max} \rangle$ (left panel): 
cold disk (blue), warm disk (green), halo (cyan), and bulge (red). The criteria adopted for this 
classification are marked with dashed lines. In the right two panels, we overlay the contours of 
these four kinds of structures on  the map of number counts using the same color. The contours at levels 
0.2, 0.4, and 0.6 are shown.}
\label{fig:intrin}
\end{center}
\end{figure*}

The criterion $C_{\rm BIC}$ is the only parameter that needs to be chosen artificially
when auto-GMM is used. A proper $C_{\rm BIC}$ can be inferred statistically from 
the large sample of galaxies in \IllustrisTNG.
To ensure that the galaxies have meaningful, well-resolved structures, we only 
use galaxies with stellar masses that exceed $10^{10}\, M_\odot$, which 
corresponds to $>10^4$ stellar particles. For each star, we specify the 
parameter $\kappa_{\rm rot} = v_\phi^2/v^2$, which measures the relative 
importance of its kinetic energy in ordered rotation. Then, the average value 
of this quantity for each galaxy, which gives an indication of its morphology 
and kinematics, is $K_{\rm rot}=\sum_i{m_i \kappa_{i, \rm rot}/M_\star}$ 
\citep{Sales2010}, where $m_i$ represents the mass of particle $i$ and 
$M_\star$ is the total stellar mass of the system. More massive galaxies 
become increasingly dominated by random motions, such that $K_{\rm rot} 
\approx 0.3$ for $M_\star \gtrsim 10^{11}\, M_\odot$ (Fig.~\ref{fig:Krot}; top 
panel). The mass ratio of spheroids $f_{\rm sph}$, estimated by summing up 
stars with $\kappa_{\rm rot}<0.5$, likely increases with increasing $M_\star$
(Fig.~\ref{fig:Krot}; bottom panel). Both $K_{\rm rot}$ and $f_{\rm sph}$ are 
kinematic indicators of the morphology of the galaxies. All the parameters 
above are calculated using the stars of radius $<30$ kpc. In Figure~\ref{fig:Krot}, 
all galaxies of stellar mass $\geq 10^{10}\, M_\odot$ are included, but
only unbarred galaxies satisfying $K_{\rm rot} \geq 0.5$ are selected for 
inferring the $C_{\rm BIC}$ of disk galaxies. We regard $K_{\rm rot} = 0.5$ 
as the criterion to separate elliptical and disk galaxies.

A massive, long bar complicates the kinematic decomposition (see 
discussion in \refsec{sec:bar}). D. Zhao et al. (2019, in preparation) find 
that a significant fraction of local disk galaxies in the TNG100 have formed a bar. 
We only focus on unbarred galaxies here, in order to obtain a clean result. 
The sample of unbarred galaxies is selected using their maximum ellipticity 
obtained from isophotal analysis of face-on images. Following standard 
convention \citep[e.g.,][]{Marinova&Jogee2007}, a galaxy is considered unbarred 
if the maximum ellipticity is less than 0.25. We obtain a total of 2994 unbarred disk
galaxies. This selected sample of unbarred disk 
galaxies is expected to have regular disky and spheroidal structures.

The $\Delta \widehat{\rm BIC}$ profiles of the selected galaxies are 
shown in Figure~\ref{fig:BICprofiles}. We vary the criterion $C_{\rm BIC}$ from
0.05 (blue dashed line) to 0.15 (red dashed line); the corresponding 
distribution of $n_c$ obtained with each $C_{\rm BIC}$ is shown in Figure~\ref{fig:nchist}. It is apparent that $C_{\rm BIC}=0.05$ gives an 
unreasonable number of components ($n_c \approx 8-11$) for most
galaxies. Both $C_{\rm BIC}=0.1$ and 0.15 yield a reasonable number of 
components ($n_c \approx 4-8$). In general, $C_{\rm BIC}=0.1$ results in 
one or two more components compared to $C_{\rm BIC}=0.15$. 

\subsection{Intrinsic Structures Found in Unbarred Disk Galaxies from \IllustrisTNG\ }
\label{sec:intri}

A large library of GMM components in unbarred disk galaxies is built up by 
applying auto-GMM to TNG100. Then we need to associate these components 
to structures with which we are familiar from observations. Visual classification is not feasible for 
a large sample of galaxies. One reasonable way to classify GMM components 
automatically is by setting appropriate criteria on the mean values of 
$j_z/j_c$, $j_p/j_c$, and $e/|e|_{\rm max}$ of each component.  
The components belonging to the same structure 
should have similar properties, and hence should also cluster in kinematic phase space.
Here we define the kinematic phase space of $\langle j_z/j_c \rangle$, $\langle j_p/j_c \rangle$, 
and $\langle e/|e|_{\rm max} \rangle$ as the mass-weighted mean values 
of $j_z/j_c$, $j_p/j_c$, and $e/|e|_{\rm max}$, respectively, of each Gaussian component. 

The 2D histogram of mean circularity $\langle j_z/j_c \rangle$ versus 
mean rescaled energy $\langle e/|e|_{\rm max} \rangle$ of all components 
of the unbarred disk galaxies is shown in the left panel of Figure~\ref{fig:intrin}. 
There are four clear, distinguishable clusters that are likely to 
correspond to intrinsic structures. We can easily classify the components 
into spheroidal and disky structures by setting a threshold circularity criterion 
$\langle j_z/j_c \rangle=0.5$ (thick dashed line). The spheroidal 
components can be classified further into bulges and halos by the criterion 
$\langle e/|e|_{\rm max} \rangle=-0.75$ (horizontal dashed line), while the 
disky components can be classified into cold disks and warm disks 
by $\langle j_z/j_c \rangle=0.85$. Here $j_p/j_c$ is not used in the 
classification, as it generally has quite a broad distribution for 
spheroids. To some extent, this is reasonable because spheroidal 
components may be composed of stars moving on highly radial orbits 
and in misaligned rotating orbits. The above strategy directly uses the data 
to statistically classify the components found by auto-GMM.

The middle and right panels of Figure~\ref{fig:intrin} plot the contours of the number distribution of the components. The four series of countours correspond to the four kinds of structures.  Both the cold and warm disks also cluster well in  $\langle j_p/j_c \rangle$, while halos and bulges have two sub-groups of $\langle j_p/j_c \rangle \approx 0.5$ and $\sim 0.2$, respectively. For spheroids, the components with $\langle j_p/j_c \rangle \approx 0.2$ are dominated by radial motions, while those with $\langle j_p/j_c \rangle \approx 0.5$ have significant, but misaligned, rotation. 

We have demonstrated that the statistical results can be used to objectively infer the intrinsic structures of galaxies.  The application of auto-GMM not only can decompose galaxies but also classify components in a completely automatic way. A few illustrative examples are shown in the next section.


\begin{figure*}[htbp]
\begin{center}
\includegraphics[width=1.0\textwidth]{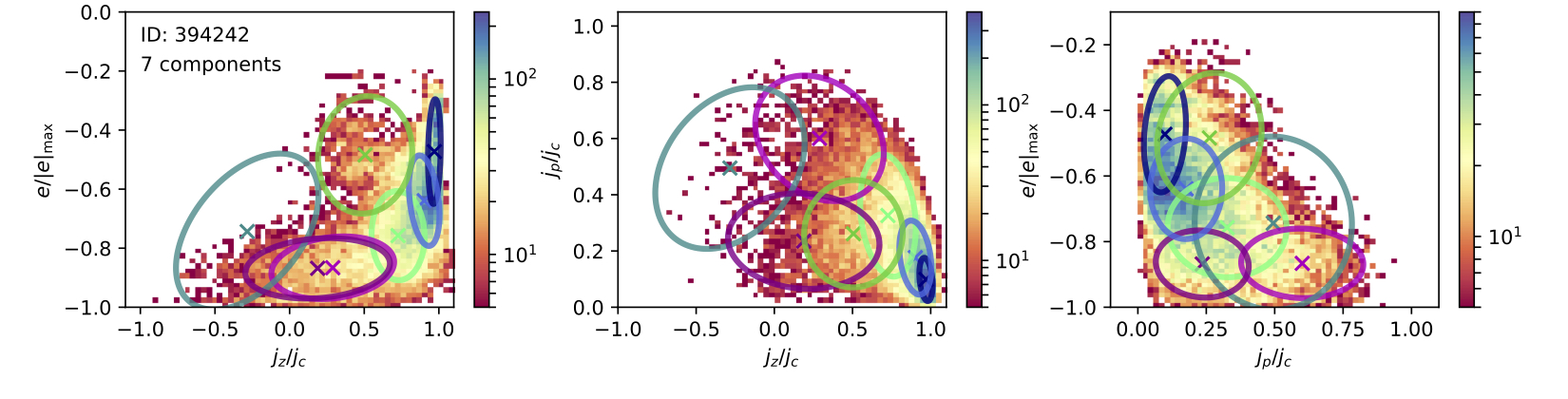}
\includegraphics[width=1.0\textwidth]{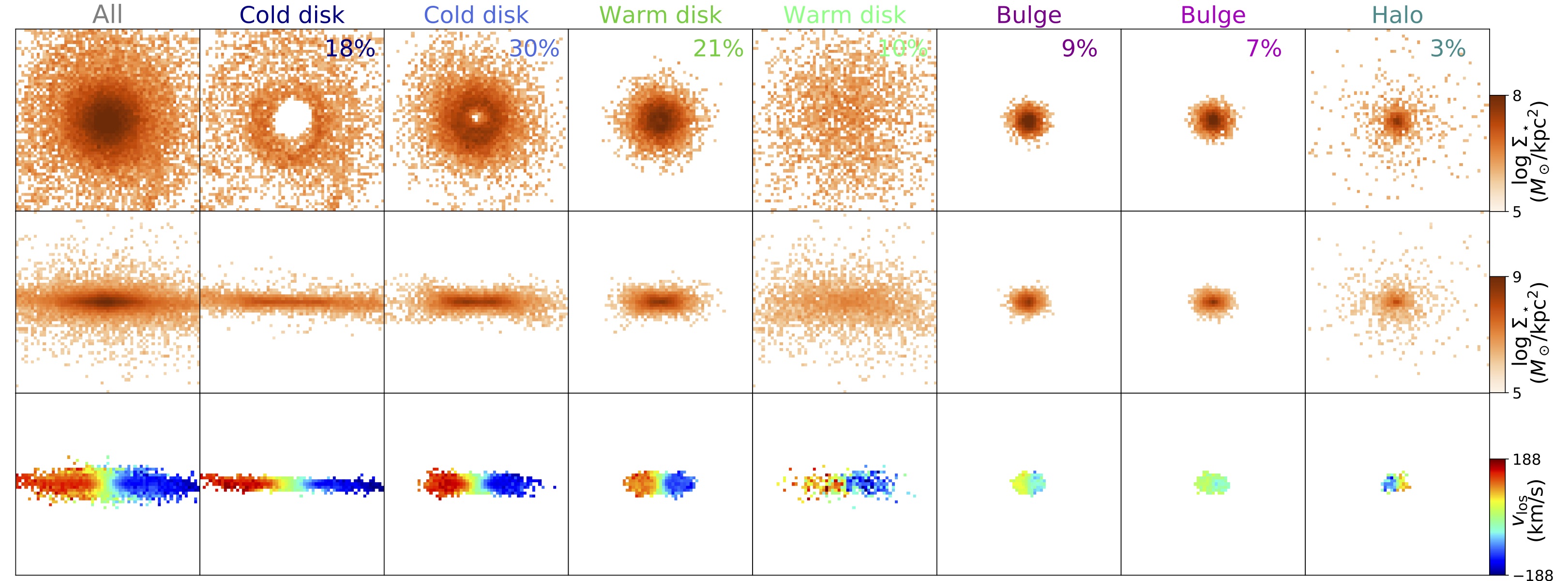}
\caption{Model D1. The top row shows the kinematic phase space of $j_z/j_c$, 
$j_p/j_c$, and $e/|e|_{\rm max}$ and the Gaussian components found using 
auto-GMM. The TNG100 ID and the number of components are labelled in the first panel. 
The log-scale color bars of the top panels show the number of stars per bin in 
phase space. Seven components are found by auto-GMM using $C_{\rm BIC}=0.1$. 
Their $63\%$ confidence ellipses are overlaid, whose corresponding means are
marked with crosses. The bottom three rows show the face-on and edge-on surface
density distributions and the edge-on line-of-sight velocity distribution, 
respectively, of each component. These components are titled according to 
visual classification, and their corresponding mass fractions are labelled, 
using the same colors as those of the ellipses in the top row. For the 
line-of-sight velocity distribution, only bins having more than five particles 
are shown. The dimensions of the $x$ and $y$ axes are 60 kpc.}
\label{fig:S1nc7}
\end{center}
\end{figure*}

\begin{figure*}[htbp]
\begin{center}
\includegraphics[width=1.0\textwidth]{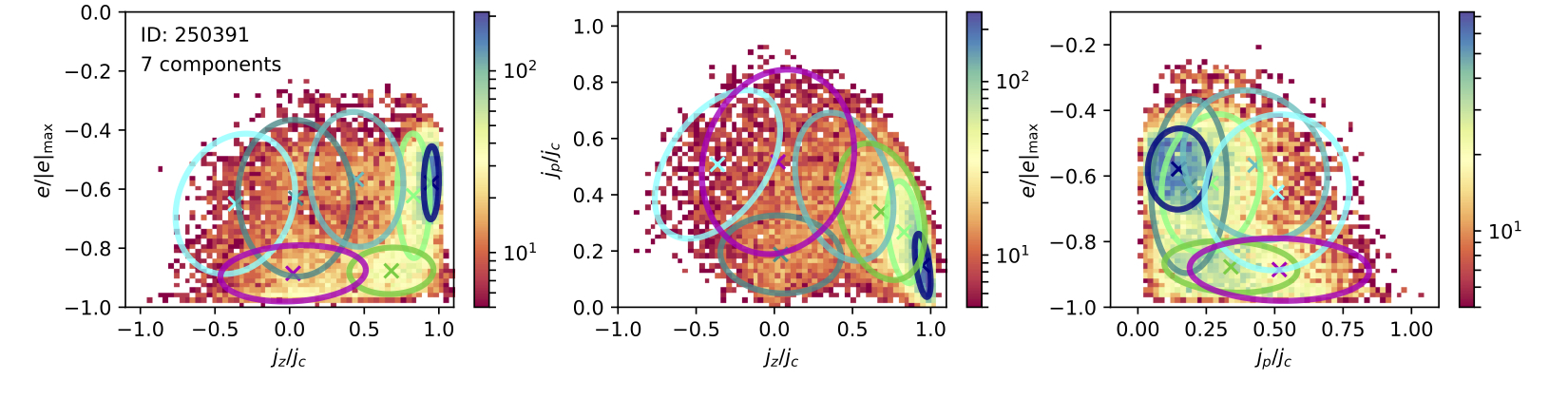}
\includegraphics[width=1.0\textwidth]{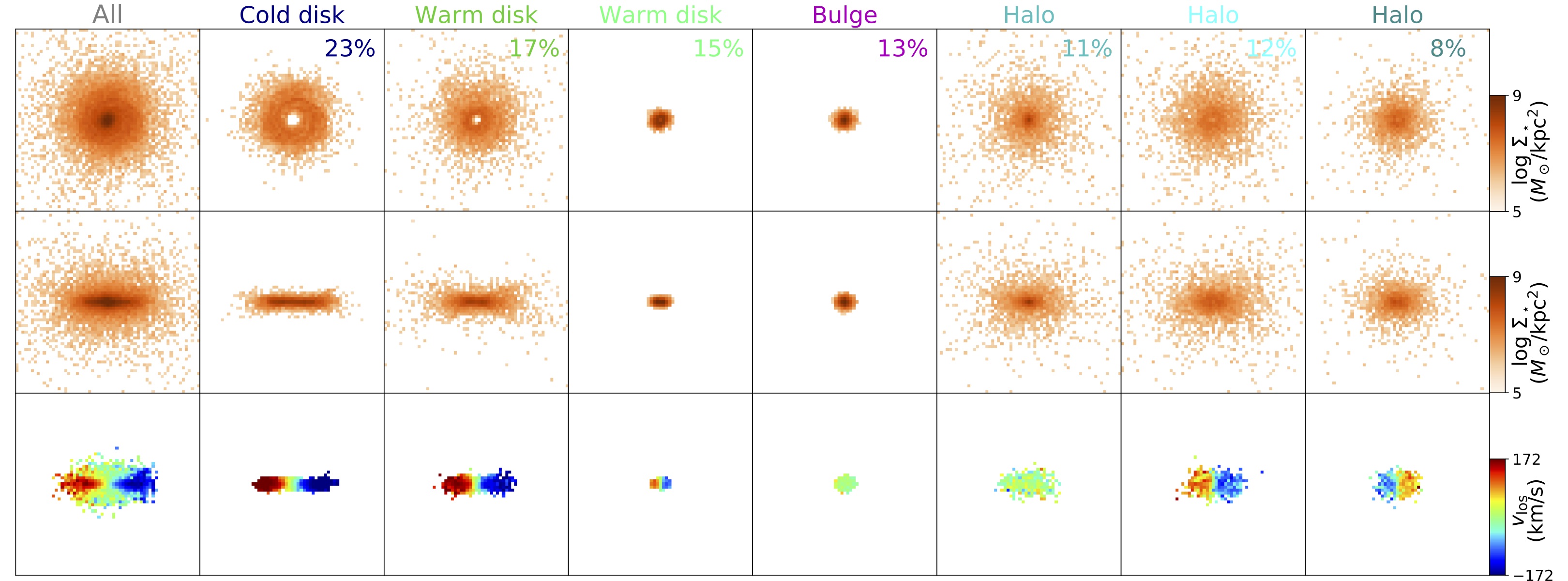}
\caption{Model D2. Seven components are found by auto-GMM. The figure uses the same conventions as Figure~\ref{fig:S1nc7}.}
\label{fig:S2}
\end{center}
\end{figure*}

\begin{figure*}[htbp]
\begin{center}
\includegraphics[width=1.0\textwidth]{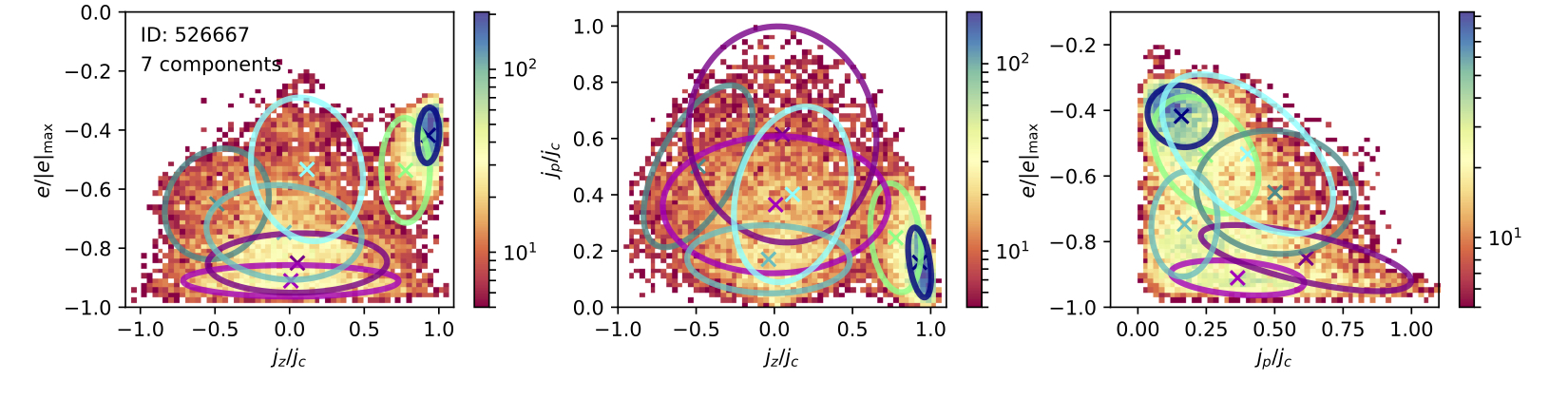}
\includegraphics[width=1.0\textwidth]{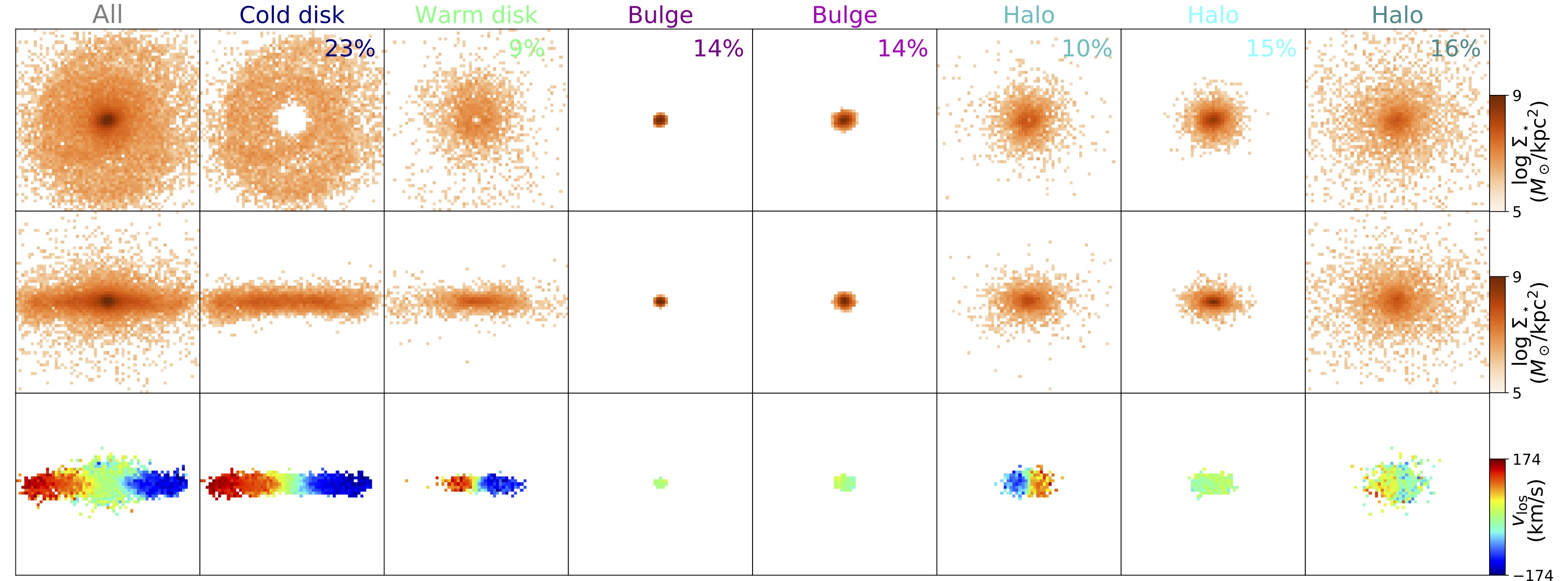}
\caption{Model D3. Seven components are found by auto-GMM. The figure uses the same conventions as Figure~\ref{fig:S1nc7}.}
\label{fig:S3}
\end{center}
\end{figure*}

\begin{figure*}[htbp]
\begin{center}
\includegraphics[width=1.0\textwidth]{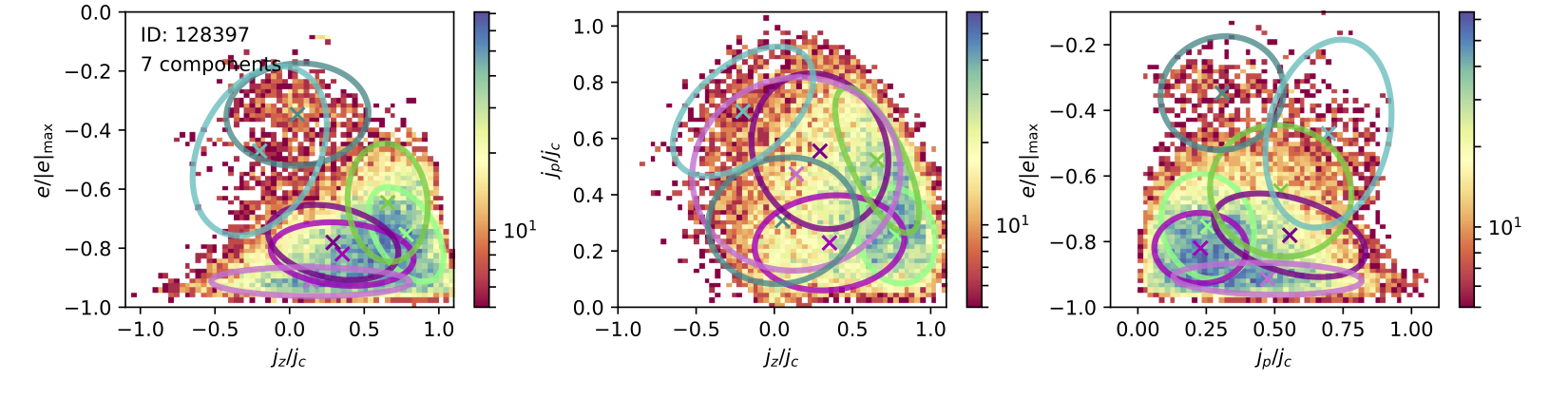}
\includegraphics[width=1.0\textwidth]{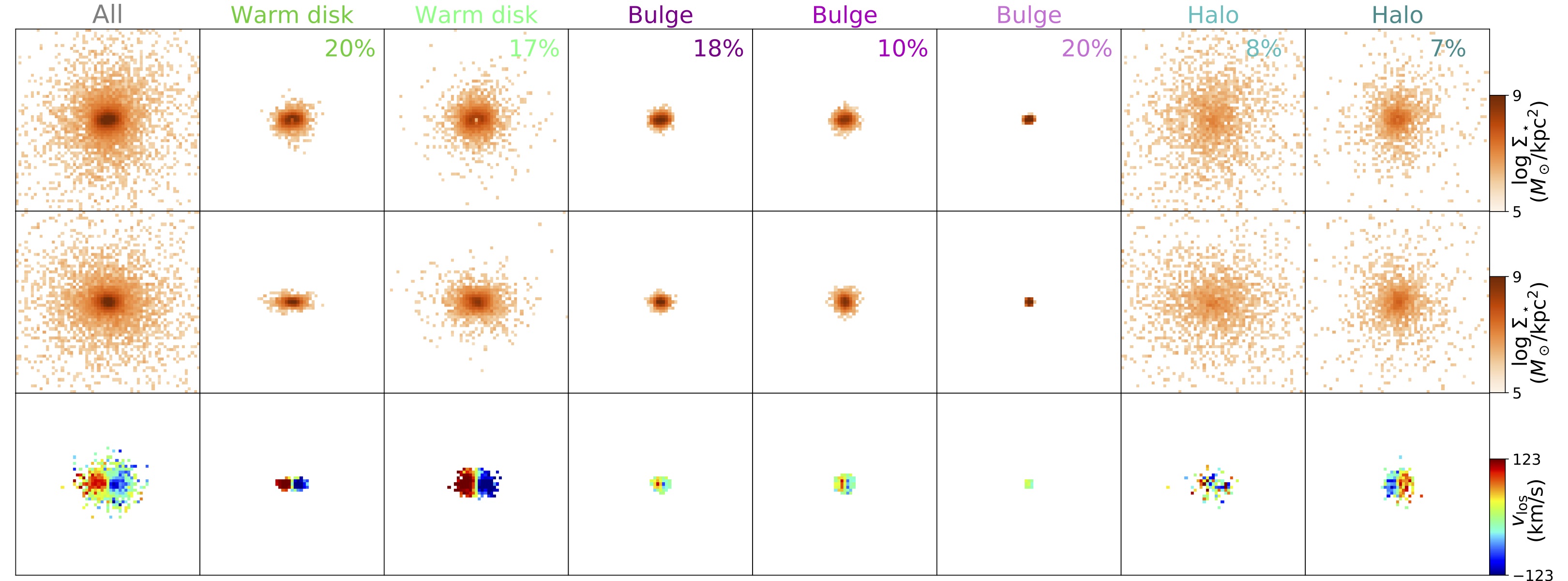}
\caption{Model E1. Seven components are found by auto-GMM. The figure uses the same conventions as Figure~\ref{fig:S1nc7}.}
\label{fig:E}
\end{center}
\end{figure*}

\begin{figure*}[htbp]
\begin{center}
\includegraphics[width=1.0\textwidth]{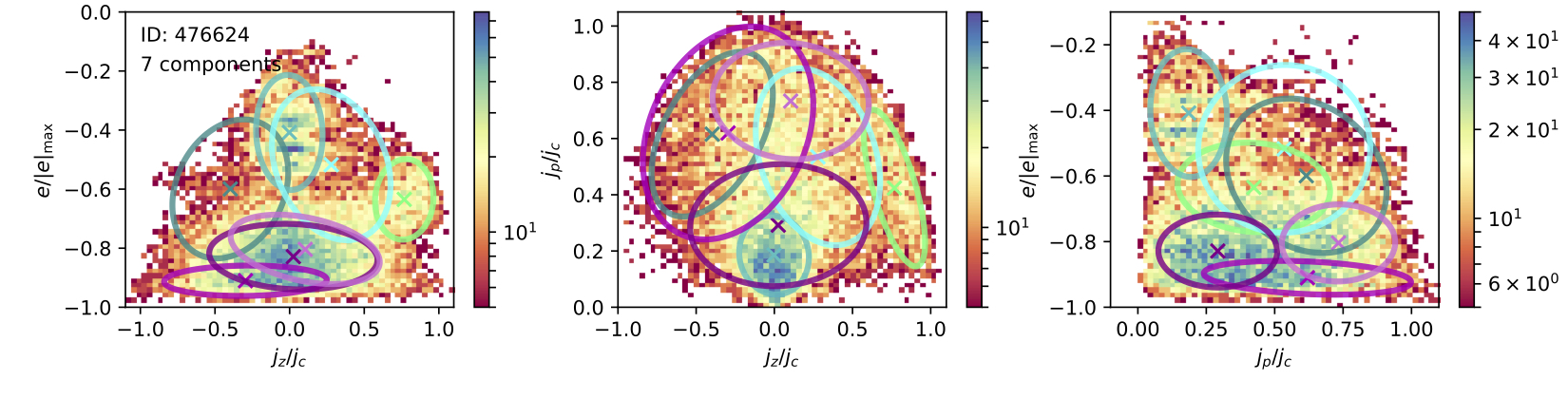}
\includegraphics[width=1.0\textwidth]{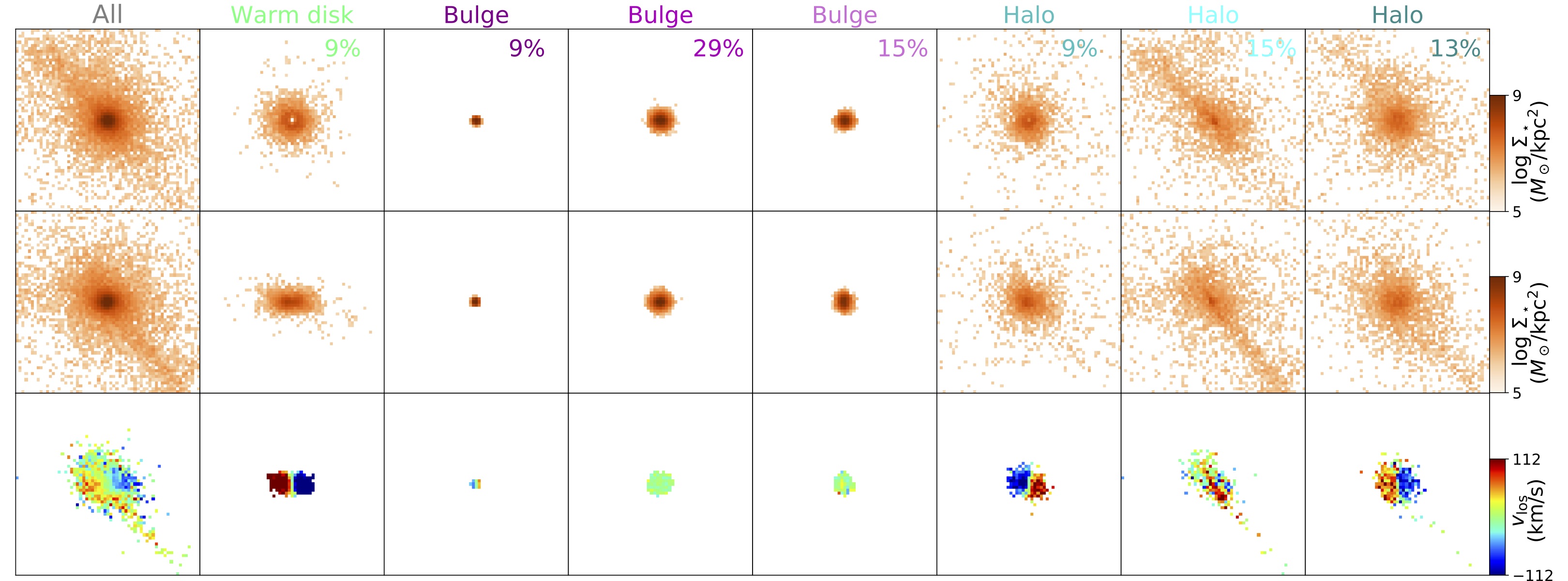}
\caption{Model E2. Seven components are found by auto-GMM. The figure uses the same conventions as Figure~\ref{fig:S1nc7}.}
\label{fig:E2}
\end{center}
\end{figure*}

\subsection{Examples of Auto-GMM Fits}

We choose a few prototype galaxies with diverse morphological and kinematic 
properties to test the performance of auto-GMM. In all cases, we adopt 
$C_{\rm BIC}=0.1$. The five prototypes---three disk galaxies (D1, D2, D3) and two 
ellipticals (E1, E2)---all have the same stellar mass ($10^{10.5}\, M_\odot$) but 
cover a range of $K_{\rm rot}$ ($\sim 0.35-0.7$) and $f_{\rm sph}$ 
($\sim 0.25-0.75$). 

Figures \ref{fig:S1nc7}-\ref{fig:E2} show the fits for models D1, D2, 
D3, E1, and E2, respectively. The first row shows diagnostic 
plots of $j_z/j_c$, $j_p/j_c$, and $e/|e|_{\rm max}$, with 63\% confidence 
ellipses of all Gaussian components overlaid. The crosses mark their means. 
The 3D kinematic phase space of the five prototypes are well fit. From the 
second to the fourth row, we show, respectively, the face-on surface density, 
the edge-on surface density, and the line-of-sight velocity distribution for 
the edge-on view.
Based on their properties, we classify the best-fit components into the
structural families presented in \refsec{sec:intri}: cold disk, warm disk, bulge, and halo.
Their corresponding mass fractions are also given. Note that 
each identified structure can contain more than one component (e.g., two cold 
disk components in D1; two bulge components in D3), and we do not ascribe any 
particular interpretation to the physical nature of such substructures here. 

Models D1 (Fig.~\ref{fig:S1nc7}) and D2 (Fig.~\ref{fig:S2}) are largely 
dominated by disky structures. Components likely associated with
cold disks and warm disks contribute 79\% to the total stellar 
mass of D1 and 55\% to the total stellar mass of D2. Only a small fraction 
of the mass in D1 arises from spheroidal components that we attribute to a 
bulge and halo. The two distinct ``cold disk'' components seen in 
the $j_z/j_c$ versus $e/|e|_{\rm max}$ plot might share the same origin, with 
one portion being slightly dynamically hotter than the other, or they might 
originate from different gas accretion events. The two substructures of the 
bulge in D1 have similar compactness and weak rotation, differing 
principally only in their non-azimuthal angular momentum $j_p$. Such a 
difference is ignored in our classification, and they are considered as 
substructures of the same bulge. 

Model D3 clearly has much more massive spheroidal components 
(\reffig{fig:S3}). Its disky components, including a cold (23\%) and a warm 
(9\%) disk, contribute only about one-third of the total stellar mass. 
The spheroidal components are impressively prominent. A clear pattern
of rotation is still evident in the kinematic phase plot of D3. 
By contrast, the diagram of $j_z/j_c$ versus $e/|e|_{\rm max}$ for model E1 
is much more irregular and exhibits far fewer meaningful features (Fig.~\ref{fig:E}).
Violent mergers may have erased much of the substructure.
Some mild rotation still exists, with $K_{\rm rot} \simeq 0.45$, arising 
mostly from an intermediate-scale disky structure. Model E1 resembles 
moderate-mass ellipticals, which typically have disky isophotes and moderate
rotation \citep[e.g.,][]{Kormendy2009}. Note that models E1 and D3 actually 
have similar rotation. However, D3 still maintains a clear disky morphology while 
E1 is quite spheroidal. E1 might be regarded as a lenticular galaxy given its
mild rotation. 

Cases with even lower values of $K_{\rm rot}$ are largely dominated by 
random motions. Model E2 is a typical elliptical galaxy with extremely weak rotation 
($K_{\rm rot} \simeq 0.35$). The pattern of E2's kinematic phase space is more regular than that of E1. 
The bulges and halos classified by the criteria from the sample of unbarred disk galaxies 
correspond to a compact nuclear component and a diffuse envelope, respectively. A tidal tail due 
to a recent minor merger is still visible in the halo. The differences between E1 and E2 indicate that they
may have experienced different assembly histories. Auto-GMM is also able to 
decompose typical elliptical galaxies, such as E2. However, many elliptical 
galaxies have a featureless kinematic phase space (e.g., E1). There is no proper way to model a 
featureless distribution even with multiple Gaussians. Thus, physically meaningless multiple Gaussians 
will be used to recover the data. Care is required in applying the auto-GMM method to elliptical galaxies.

It is worth emphasizing that the relation between the structures decomposed by kinematics and those 
from morphological observations is still unclear. The morphologies of the structures found 
by auto-GMM here are roughly consistent with our expectations of thin disks, thick disks, bulges, and halos. 
However, there are some essential differences. On the one hand, the bulge defined in 
kinematics is the tightly bound/compact part of spheroids, while the halo is the diffuse part. Halos 
do contribute to the central density, which is indistinguishable in observations of most of external galaxies. 
Thus, the inner part of kinematic halos will be considered as part of (classical) bulges in observations. 
Whether bulges and halos are formed in the same way, namely through mergers, is beyond the scope of 
this paper. On the other hand, warm disks may be related to thick disks and pseudo bulges 
in observations, and may have formed via very diverse pathways. 
Forthcoming papers will statistically investigate the properties and evolution of the structures identified here.

\subsection{The Failure of Auto-GMM Fits in Barred Galaxies}
\label{sec:bar}

\begin{figure*}[htbp]
\begin{center}
\includegraphics[width=1.0\textwidth]{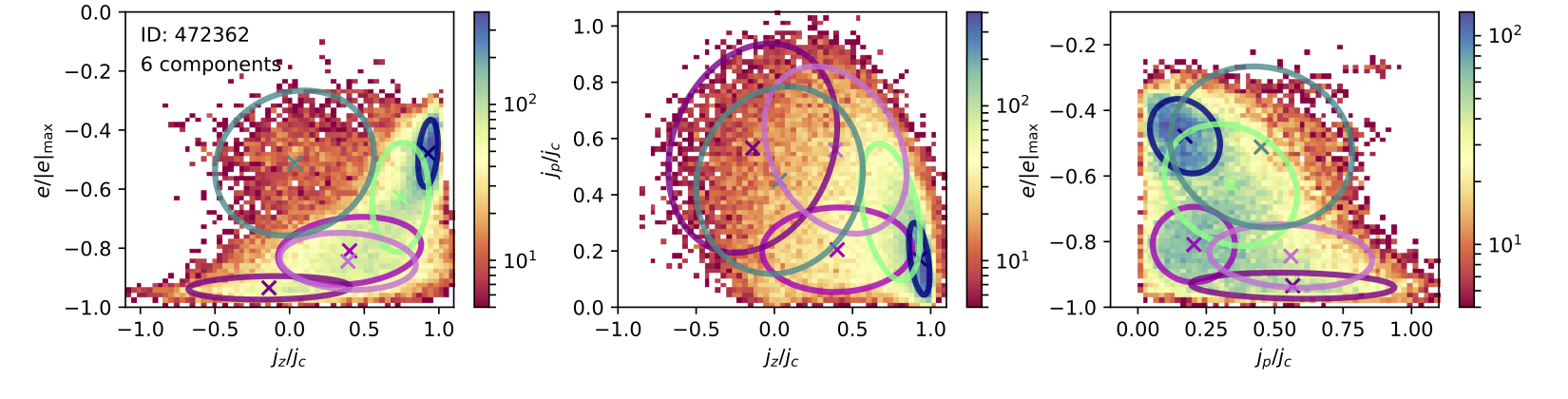}
\includegraphics[width=0.95\textwidth]{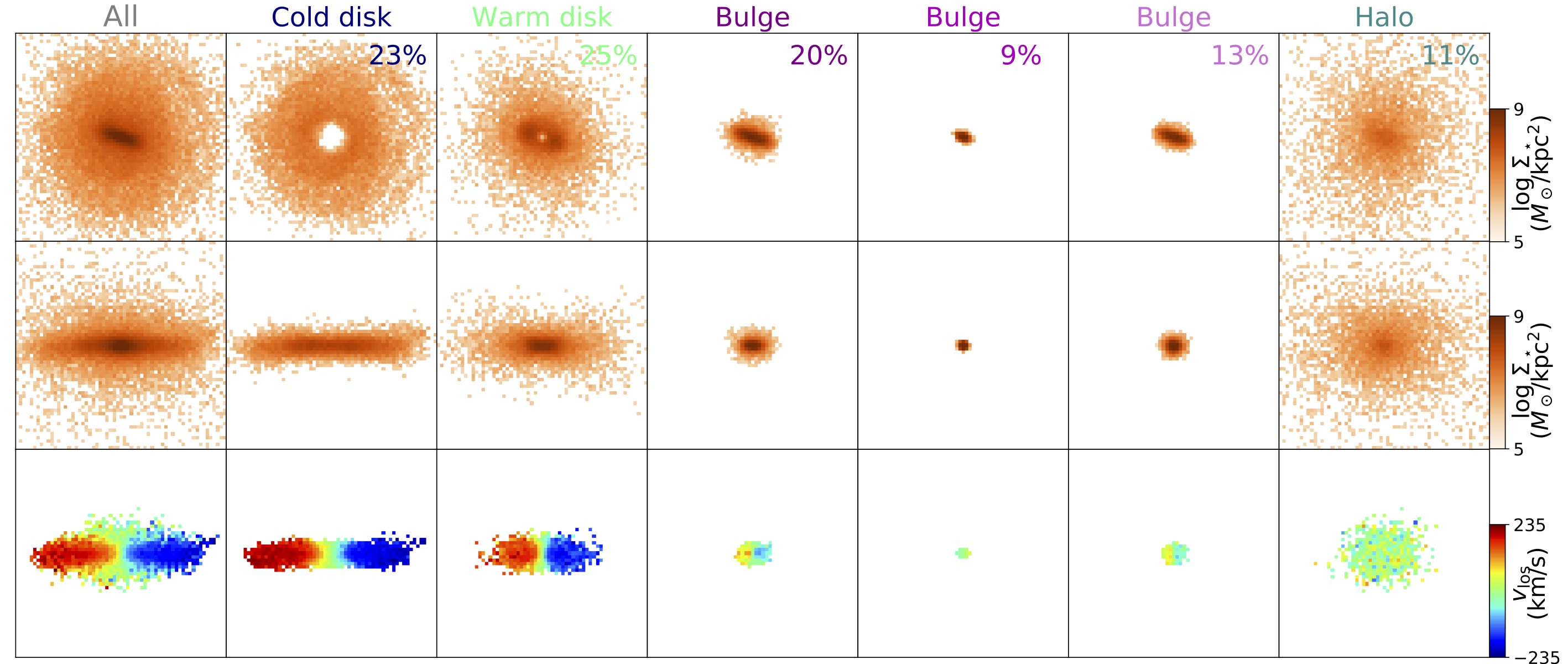}
\caption{A galaxy hosting a strong bar. Six components are found by auto-GMM with $C_{\rm BIC}=0.1$. The figure uses the same conventions as Figure~\ref{fig:S1nc7}.}
\label{fig:bar}
\end{center}
\end{figure*}

Particles moving on bar orbits have complex kinematics that are unlikely to be well 
described by the phase space of $j_z/j_c$, $j_p/j_c$, and $e/|e|_{\rm max}$. 
Figure~\ref{fig:bar} shows an example of an auto-GMM fit of a typical barred galaxy 
from \IllustrisTNG.  At a given radius, particles moving on bar orbits rotate 
more slowly compared with those on circular orbits, and $j_z/j_c$ decreases
gradually with decreasing $e/|e|_{\rm max}$. As a consequence, bar particles 
significantly pollute the components having moderate rotation, such as warm 
disks. At the same time, bar particles with $j_z/j_c<0.3$ may also
influence significantly the kinematic decomposition of slowly rotating 
components, probably even bulges. Under this circumstance, the mass of the bulge is clearly 
overestimated. Bar particles do not cluster well in this kinematic phase space, 
as shown in the first row of Figure~\ref{fig:bar}. They instead drive significant 
mixture in the kinematic phase space between disks and spheroids. Therefore, the 
auto-GMM method fails to reliably decompose barred galaxies. 

\section{Summary}
\label{summary}

We have described an automated method, auto-GMM, that generalizes the Gaussian 
mixture models to decompose the stellar kinematics of simulated unbarred galaxies. A 
modified version of the Bayesian information criterion is used to 
infer the optimal number of statistically significant Gaussian components 
to fit the data. 

We demonstrate that the simulated galaxies display rich substructures that can
be identified and decomposed effectively by auto-GMM in the kinematic phase space of 
the stellar particles. Each substructure is a 3D Gaussian component. 
The substructures belonging to the same structure also cluster 
in the diagram of the mean circularity versus the compactness (rescaled energy) of the 
Gaussian components. Taking advantage of a large sample of galaxies in the cosmological 
simulation \IllustrisTNG, four kinds of intrinsic structures are identified: 
cold disks, warm disks, bulges, and halos. 
While the present study does not ascribe any rigorous 
physical interpretation to the decomposed individual components, we illustrate 
the power of the auto-GMM method to isolate features that can be plausibly 
associated with morphological components (cold disk, warm disk, bulge, halo) 
traditionally associated with structures in the Hubble sequence of galaxies.

Our proposed method is automated, fast, and effective. It is a powerful tool 
to analyse a large data set of galaxies from cosmological simulations to gain 
insights into the origin and nature of galaxy structure. 
In forthcoming work, we will statistically investigate the properties of structures in thousands of 
galaxies from IllutrisTNG. We hope that the results can help interpret observations and provide more insight 
into the formation and evolution of real galaxies.

\begin{acknowledgements}
This work was supported by the National Science Foundation of China (11721303) and the National Key R\&D Program of China (2016YFA0400702).   M.D. is also supported by the grants ``National Postdoctoral Program for Innovative Talents'' (\#8201400810) and ``Postdoctoral Science Foundation of China'' (\#8201400927) from the China Postdoctoral Science Foundation. D.Y.Z. and J.S. acknowledge the support by the Peking University Boya Fellowship. V.P.D. was supported by STFC Consolidated grant ST/R000786/1. The TNG100 simulation used in this work, one of the flagship runs of the IllustrisTNG project, has been run on the HazelHen Cray XC40-system at the High Performance Computing Center Stuttgart as part of project GCS-ILLU of the Gauss centres for Supercomputing (GCS). The authors thanks all the IllustrisTNG team for making the IllustrisTNG data available to us prior to the public release. We thank the anonymous referee for valuable comments. We also thank Dandan Xu and Aura Obreja for constructive discussions. This work is highly supported by the High-performance Computing Platform of Peking University, China. The analysis was performed using \texttt{Pynbody} \citep{pynbody}.
 
\end{acknowledgements}



\bibliographystyle{apj}
\bibliography{Reference_lib}

\end{document}